# Text Mining System For Non- Expert Miners

**Ramya P**
**PG Scholar**
**SSN College of Engineering**
**Chennai, India**
**ramyaappaya@gmail.com**

**Sasirekha S**
**Assistant Professor**
**SSN College of Engineering**
**Chennai, India**
**sasirekhas@ssn.edu.in**

***Abstract***-- Service oriented architecture integrated with text mining allows services to extract information in a well defined manner. In this paper, it is proposed to design a knowledge extracting system for the Ocean Information Data System. Deployed ARGO floating sensors of INCOIS (Indian National Council for Ocean Information Systems) organization reflects the characteristics of ocean. This is forwarded to the OIDS (Ocean Information Data System). For the data received from OIDS, pre-processing techniques are applied. Pre-processing involves the header retrieval and data separation. Header information is used to identify the region of sensor, whereas data is used in the analysis process of Ocean Information System. Analyzed data is segmented based on the region, by the header value. Mining technique and composition principle is applied on the segments for further analysis.

## I. INTRODUCTION

Text mining is described as the process of deriving high quality information from text. High quality information is derived from the patterns and trends. Text mining involves the processes of structuring the input text, deriving patterns from the structured data and finally the interpretation and evaluation of output.

View of text mining is an extension of data mining or knowledge from structured databases. Knowledge discovery from textual database refers generally to the process of extracting interesting or non-retrieval patterns or knowledge from unstructured text documents.

In today's competitive market, information is one of the main managerial assets since its analysis helps in effective steering. The concept of text mining is used in E-learning Web Miner Application [1]. It is a graphical user interface built with several operational modes for its users. For each mode of operation separate template was designed using Java.

Data mining technique in attrition analysis [2] is to identify a group of customers who have a high probability to attrite, and then the company can conduct marketing campaigns to change the behaviour in the desired direction. Nowadays the majority of large companies and corporations have to a greater or lesser extent a Data Warehouse and they use OLAP tools to extract and analyze the information which allows them to stand themselves in the market.

However, although there are areas where data mining techniques are being used more and more, such as business [3], marketing, education, banking, health and security systems, and so on , their use is still not generalized. This is mainly due to the fact that data mining projects need highly qualified professionals (expert data miners) to achieve, in reasonable time, useful results for business. One of the reasons for which expert data miners are required is that the knowledge discovery in databases (KDD) process involves multiple stages, and regretfully, in each one, there are a large number of decisions that have to be taken.

Data mining have induced interest [4] among the business community, particularly in large corporations with strong collection of data about their customers and business operations. Increased concentration towards business applications has necessitated even more requirements for knowledge discovery projects.

Extraction of information from the unstructured text [5], which will help all kinds of users. Information extraction typically is performed in the form of analysis pipelines. The pipeline stages are formant conversion, sentence splitting, tokenization, word stemming and annotation of tokens.

Mining technique in decision support system [6] is described with the temporal data. In this relationships between the events that affect the decision are discussed. This relationship is defined using the unsupervised learning technique of data mining. It not just defines the relationships between events, also extracts the interesting patterns and boundaries present in the system. The two methods of mining technique are supervised learning and unsupervised learning technique. Supervised learning technique is used in predictive statistical techniques whereas unsupervised learning method does not use dependent variables. It searches for the patterns and events.

This paper is organized in the following sections. Section 2 involves the system description and section 3 describes proposed work of the system. Section 4 gives the details of simulation results and finally Section 5 projects the conclusion.

## II. SYSTEM DESCRIPTION

A system modelled to service all community is introduced in this paper. The concepts used to achieve this system are text mining and service oriented architecture. Services provided by the proposed system are fishing zone advisory and tsunami alerts. This can be performed by the analysis of ocean information. Ocean information is retrieved from the OIDS of INCOIS organisation. Data from the OIDS is pre-processed with some constraints as adding precision points to the processed data. Pre-processing is literally to minimize the difficulties in analysis process. Measurement of index value and pattern extraction becomes easier with the pre-processed data. From the measured values a graphical representation is obtained, which clearly states the system's performance.

## III. PROPOSED WORK

Processing steps discussed in this paper are as follows: Pre-processing, Analysis, Exploration and Interpretation. Pre-processing step is generally defined as the data preparation process In this paper pre-processing step is for converting the hexa-decimal data to decimal data and adding precision values to the data.

Analysis process of this system involves the calculation of ocean index value. Index value is to represent the nature and mining technique is to extract interesting patterns of the system. Formula used to calculate the index value is,

$$N = 1.3247 - 2.5 \times 10^{-6} T^2 + S(2 \times 10^{-4} - 8 \times 10^{-7} T) + \frac{3300}{P^2} - \frac{3.2 \times 10^7}{P^4} \quad (1)$$

The parameters used in the index calculation are described as N is the oscillation index, T is the ocean temperature, S is the salinity of ocean water and P is the pressure value. Application of mining technique is known as the extraction process. As we are processing the ocean data, unsupervised learning data technique is applied in the here.

After the calculation of ocean index, association rule mining algorithm (an example of unsupervised learning method) is applied. It works by calculating the confidence and support metrics of the given data. Before this calculation, given data set is segmented by the time lag into events. The support of the rule is the number of times the rule holds same event in the database whereas confidence rate of an episode is calculated as,

$$r = \chi \text{ [win a]} \Rightarrow \psi \text{ [win c]} \quad (2)$$

The parameters win a, win c denotes the window values, $\psi$, $\chi$ are time delay values which are estimated as the time difference between the similar patterns. The interpretation technique is used to deploy the extracted results. Here composition technique is used as an interpretation. Composition is the principle of service orientation and is used to compose the results of different operating systems. The analysis results of the ocean data are composed using this principle. The composition process gives us the details of fishing zone and tsunami alerts respective to the areas. This result is documented in a table format with the necessary fields, denoting the analysis process.

## IV. IMPLEMENTATION

### *A. Data Set*

INCOIS is an acronym for Indian National Centre for Ocean Information System. It is an organization of central ministry department. This organization aims to provide a system which has the current analysis of ocean characteristics.

To achieve this it deploys ARGO floating sensors in the ocean, which reads the physical characteristics like temperature, salinity and pressure. The characteristic values are transmitted to OIDS (Ocean Information Data System) through satellites. Now the data received on OIDS is forwarded for the analysis process. In fig.1 the sample input dataset is given.

```
02602 2902102 65 32 K 2 2003-01-10 11:50:18,0.691 76.559 0.000 401647210
2003-01-10 12:49:18 1 EE 05
35 9D 89 3E 07 CB
37 09 89 54 07 62
38 30 89 63 06 FF
38 E7 89 67 06 A0
2014-05-01 00:46:47 1 9F 06 3A 40
89 68 06 3A
3B 5D 89 66
05 D5 3C 38
E1 C7 9E 3C
79 C3 8A 14
F0 B9 4A 12
F3 8E 39 F0
02602 32134 73 32 K 2 2003-01-10 14:34:18 0.706 76.542 0.000 401647210
2003-01-10 14:28:18 1 4D 0B 70 B9
89 D2 00 E1
70 C9 89 D1
00 BE 70 D0
89 C9 00 A4
70 D7 89 C6
00 7C 70 E6
89 C4 00 6E
```

Fig.1 Dataset from OIDS

This dataset contains the sensor header value and data. Here data denotes the values of temperature, salinity and pressure. This data is pre-processed for the analysis process. Pre-process step involves the sensor header value extraction and data separation which is explained in Fig.2.

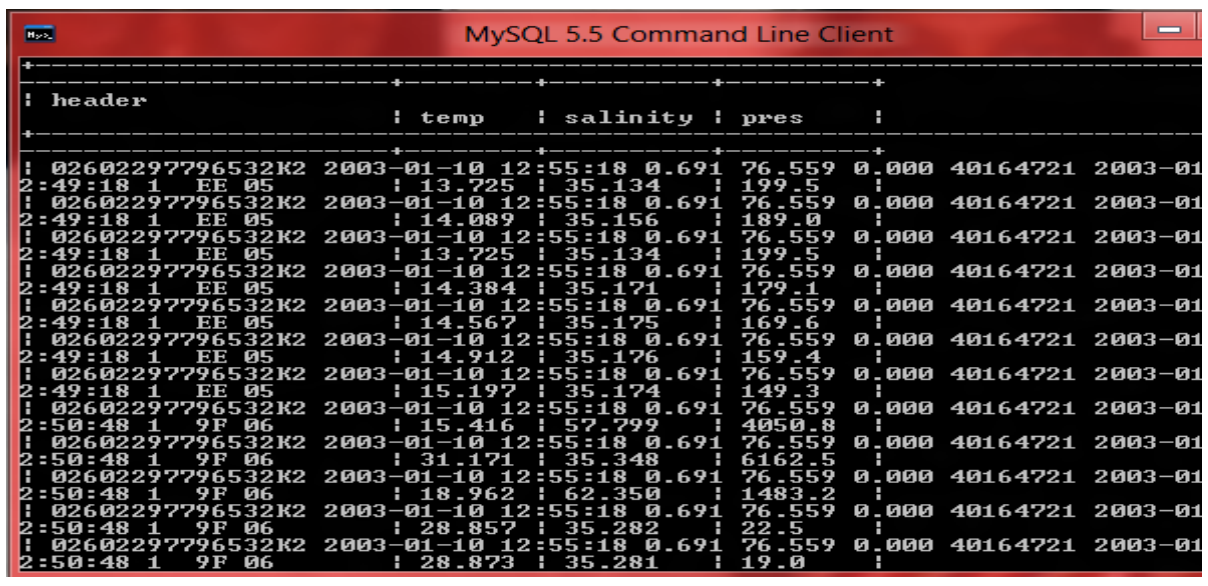

Fig.2 Separated Header and Data

Data is separated from the set and added with precision values in the pre-processing step. Also the header extracted from the set is used to distinguish the sensors deployed. After the separation data is segmented into regions using its header value. It is explained in Fig. 3.

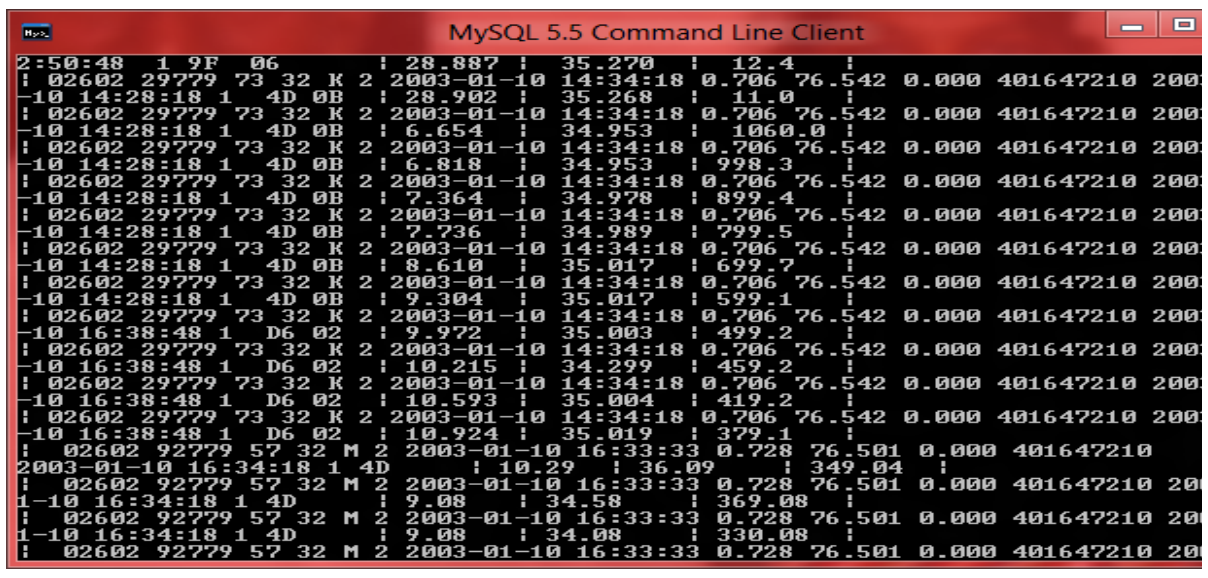

Fig.3 Region based segmentation

Data with same header values belong to a certain region are grouped in this step. Index value is estimated after the segmentation of regions and is given in Fig. 4.

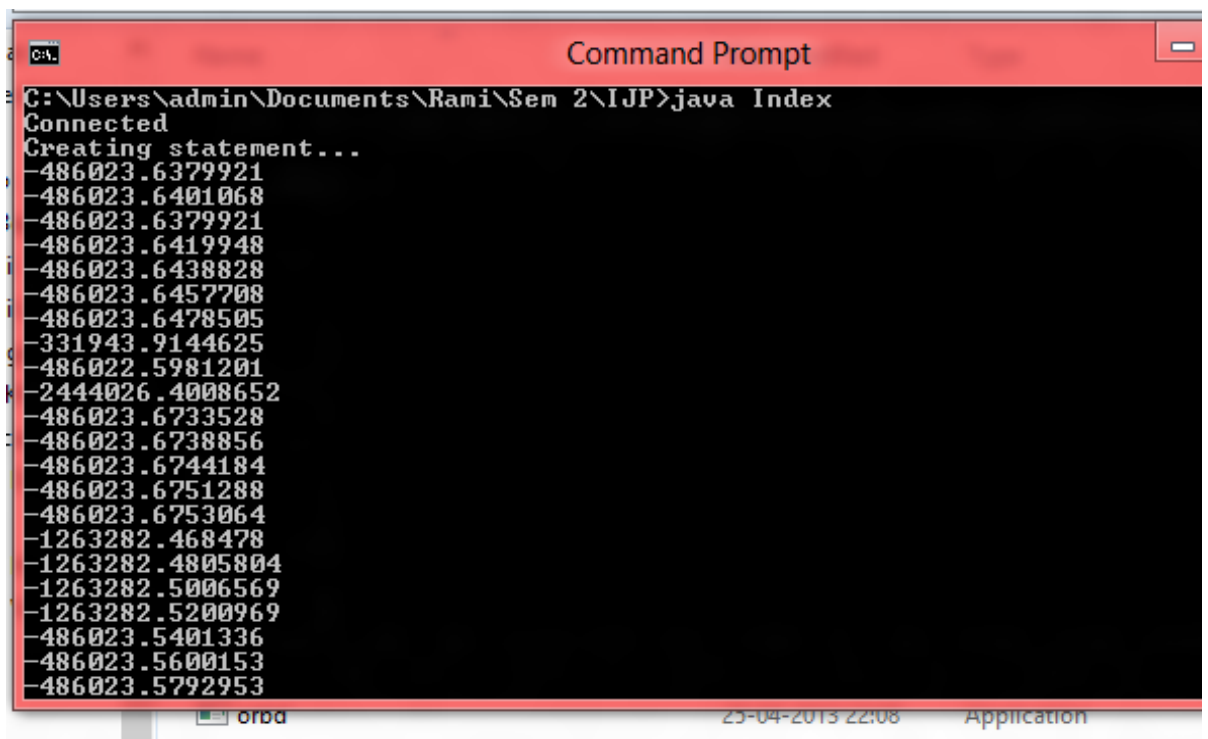

Fig .4 Ocean Index Value

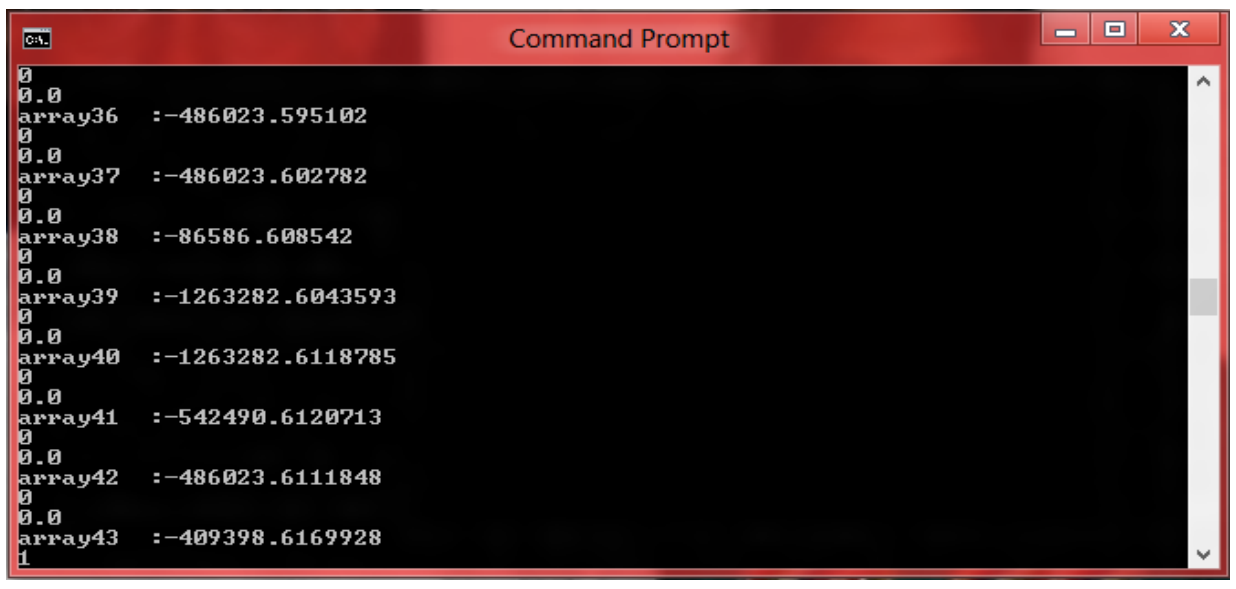

Fig.5 Support and Confidence metric

Fig.6 indicates the graphical representation of the time and index value. This plot has an average min and max values. When the index has its value lower or greater than its average min and max value, it is defined that the area might have strong waves.

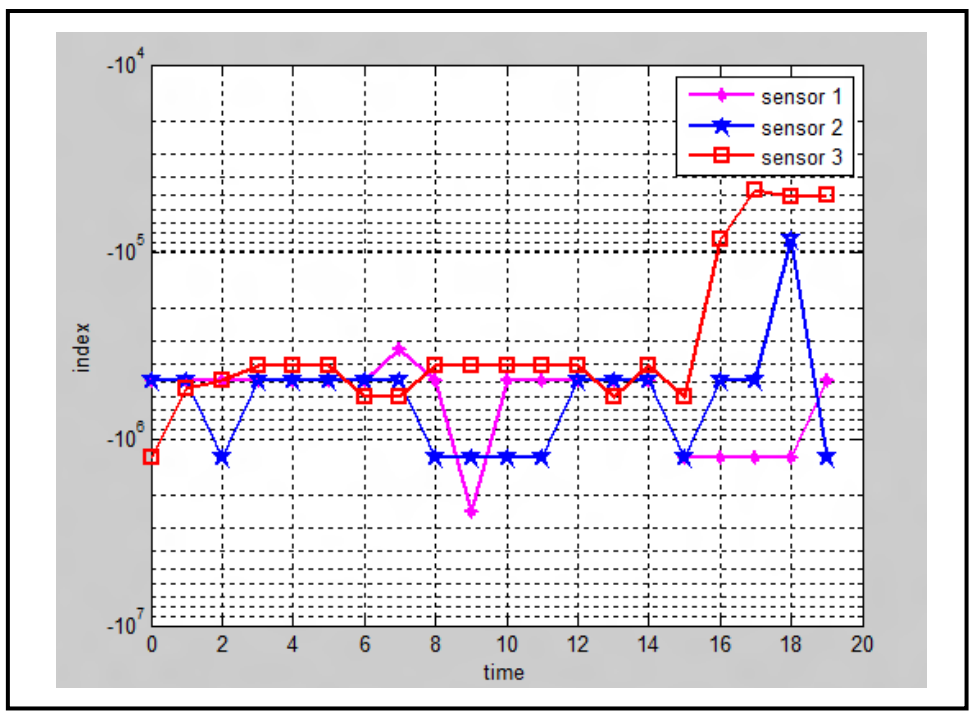

Fig.6 Analysis

Method to extract the similar pattern is depicted in fig.6. Here the value of confidence is plotted against the time value.

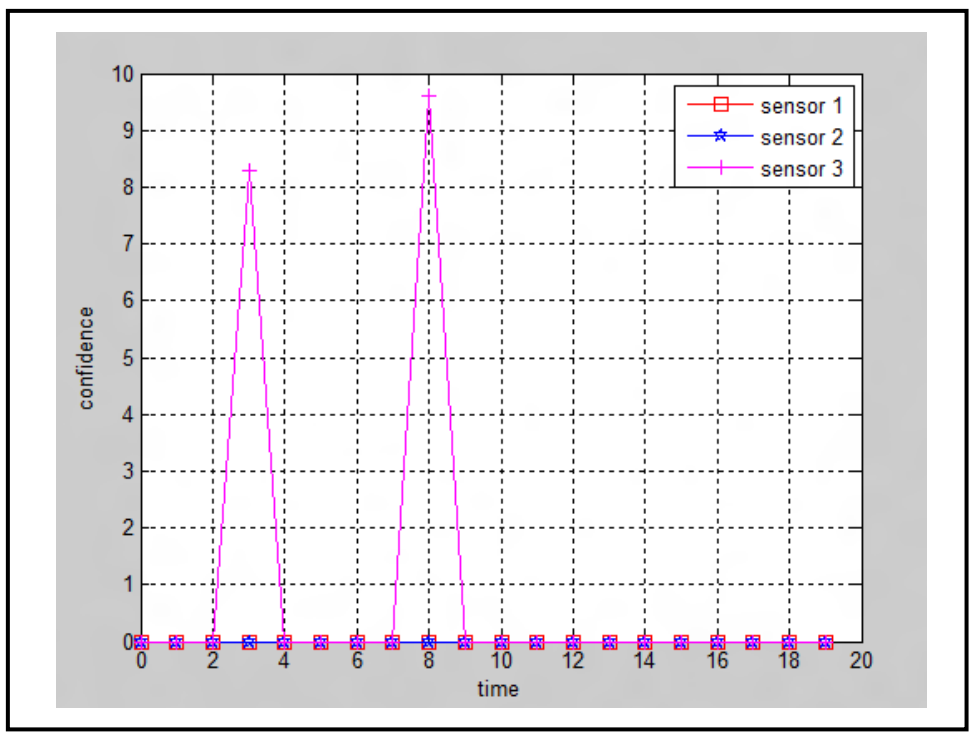

Fig.7 Time Vs Confidence

This graphical representation defines the fishing zone of the ocean. Inference obtained from the graph is that when the confidence value goes to its peak value with respect to time, then the area is referred as potential fishing zone.

## V. CONCLUSION

Text Mining is a mining technique which extracts the characteristics of any system. In this project, it is proposed to generate a report about the ocean system. This report acts as a warning system of disasters like tsunami and storm. Also its information database contains the location of ores and mines under water, which increases the economy of our country. Tourism Board and Oil, Gas Producers of our country highly depending on the ocean for their growth. It helps the fisherman by listing the major fishing zones. It also used to alert the public at the time of disasters.

## I. REFERENCES

[1] Marta Zorrilla, Diego García-Saiz (2012), “A service oriented architecture to Provide Data Mining services for Non-expert Miners,” Decision Support Systems 55, Pages: 399–411, (2013).

[2] X. Hu, “A data mining approach for retailing bank customer attrition analysis,” Applied Intelligence 22, Pages:47–60, (2005).

[3] H. Jiawei, K. Micheline, “Data mining: concepts and techniques,” Elsevier Inc.,(2006).

[4] A. Arsanjani,” Service-oriented modelling and architecture: How to identify, specify, and realize services,” (2011).

[5] M.C. Chen, A.L. Chiu, H.H. Chang, “Mining changes in customer behaviour in retail Marketing,” Expert Systems with Applications 28, Pages:773–781, (2005).

[6] S.Y. Kim, T.S. Jung, E.H. Suh, H.S. Hwang, “Customer segmentation and strategy development based on customer lifetime value: a case study,” Expert Systems with Applications 31, Pages:101–107, (2006).

[7] Randy Kerber, Hal Beck, Tej Anand,Bill Smart, “Active Templates: Comprehensive Support for the Knowledge Discovery Process “,from KDD Processings.

[8] Seyed Mohammad Seyed Hosseini, Anahita Maleki, Mohammad Reza Gholamian, “Cluster analysis using data mining approach to develop CRM methodology to assess the customer loyalty,” Expert Systems with Applications 37, Pages:5259–5264, (2010).

[9] Johannes Starlinger, Florian Lietner, Alfonso Valencia, Ulf Leser, “SOA based integration of Text Mining Services,” Congress On Services-I, (2009).

[10] Q. Yang, X. Wu, 10 challenging problems in data mining research, “International Journal of Information Technology & Decision Making,” Pages:597–604, , (2006).

[11] P.C. Wong, P. Whitney, J. Thomas, “Visualizing association rules for text mining, Proc. Of IEEE Symposium on Information Visualization,” IEEE Computer Society, Pages:120–128, (1999).

[12] Sherri K. Harms, Jitender S.Deogun, “Sequential Association Rule Mining with Time Lags, Intelligent Information Systems 22:1, 7-22,” (2004).

[13] R. Mazza, V. Dimitrova, “CourseVis: a graphical student monitoring tool for supporting instructors in web-based distance courses,” International Journal of Human-Computer Studies 65 (2007) 125–139.

[14] E.W.T. Ngai, Yong Hu, Y.H. Wong, Yijun Chen, Xin Sun, “The application of data mining techniques in financial fraud detection: a classification framework and an academic review of literature,” Decision Support System 50 (2011) 559–569.

[15] M.P. Papazoglou, W. van den Heuvel, “Service-oriented architectures: approaches, technologies and research issue,” The VLDB Journal 16 (2007) 389–415.

[16] M.E. Zorrilla, E. Álvarez, “MATEP: monitoring and analysis tool for e-Learning plat--forms,” Proc. 8th IEEE International Conference on Advanced Learning Technologies, Santander, 2008.

[17] T. Erl, “Service-Oriented Architecture: Concepts, Technology,” and Design, 1st ed.Prentice Hall, 2005.

[18] A. Haira, D. Birant, A. Kut, “Improving quality assurance in education with web-based services by data mining and mobile technologies,” Proc. of the 2008 Euro American Conference on Telematics and information Systems, ACM, New York, NY, 2009, pp. 1–7.

[19] U.M. Fayyad, G. Piatetsky-Shapiro, P. Smyth, R. Uthurusamy, “Advances in Knowledge Discovery and Data Mining Boston,” AAAI/MIT Press, 1996.

[20] H. Jiawei, K. Micheline, “Data mining: concepts and techniques,” Elsevier Inc., San Francisco, 2006.

[21] R. Hijon, A. Velázquez, “E-learning platforms analysis and development of students tracking functionality,” in: E. Pearson, P. Bohman (Eds.), Proc. of World Conference on Educational Multimedia, Hypermedia and Telecommunications, Chesapeake, 2006, pp. 2823–2828.

[22] L. Feng, J. Yu, H. Lu, J. Han, “A template model for multidimensional inter-transactional association rules,” The VLDB Journal 11 (2002) 153–175.

[23] D. Delen, “A comparative analysis of machine learning techniques for student retention management,” Decision Support Systems 49 (2010) 498–506.

[24] I. Douglas, “Measuring participation in internet supported courses,” in Proc. of the 2008 International Conference on Computer Science and Software Engineering, IEEE Computer Society, Washington, DC, 2008, pp. 714–717.

[25] Sudipto Guha, Rajeev Rastogi, and Kyuseok Shim, (1998), “ROCK: A Robust Clustering Algorithm for Categorical Attributes,” In Proceedings of the 15th International Conference on Data Engineering, 1999.

[26] C.S. Potter, S. A. Klooster, and V. Brooks, “Inter-annual variability in terrestrial net primary production: Exploration of trends and controls on regional to global scales,” Ecosystems, 2(1): 36-48 (1999).

[27] Pang-Ning Tan, Michael Steinbach, Vipin Kumar,Steven Klooster, Christopher Potter, Alicia Torregrosa, “Finding Spatio-Termporal Patterns in Earth Science Data: Goals, Issues and Results,” Temporal Data Mining Workshop, KDD2001 (2001).

[28] G. H. Taylor, “Impacts of the El Niño/Southern Oscillation on the Pacific Northwest”(1998) http://www.ocs.orst.edu/reports/enso_pnw.html.

[29] M. Steinbach, P. N. Tan, V. Kumar, C. Potter, S.Klooster, A. Torregrosa, “Clustering Earth Science Data: Goals, Issues and Results”, In Proc. of the Fourth KDD Workshop on Mining Scientific Datasets (2001).

[30] L. Ertöz, M. Steinbach, and V. Kumar, "Finding Topics in Collections of Documents: A Shared Nearest Neighbor Approach," Text Mine '01, Workshop on Text Mining, First SIAM International Conference on Data Mining, Chicago, IL, (2001).